\documentstyle[multicol,eqsecnum,aps,psfig]{revtex}
\renewcommand{\narrowtext}{\begin{multicols}{2}
\global\columnwidth20.5pc\noindent}
\renewcommand{\widetext}{\end{multicols}
\global\columnwidth42.5pc}
\begin{document}
\draft
\preprint{12 December 2000}
\title{Low-Temperature Properties of Quasi-One-Dimensional
       Molecule-Based Ferromagnets}
\author{Takashi Nakanishi, Shoji Yamamoto and
        T$\hat{\mbox o}$ru Sakai$^*$}
\address{Department of Physics, Okayama University,
         Tsushima, Okayama 700-8530, Japan\\
         $^*$Faculty of Science, Himeji Institute of Technology,
         Ako, Hyogo 678-1297, Japan}
\date{12 December 2000}
\maketitle
\begin{abstract}
Quantum and thermal behaviors of low-dimensional mixed-spin systems
are investigated with particular emphasis on the design of
molecule-based ferromagnets.
One can obtain a molecular ferromagnet by assembling molecular bricks
so as to construct a low-dimensional system with a magnetic ground
state and then coupling the chains or the layers again in a
ferromagnetic fashion.
Two of thus-constructed quasi-one-dimensional bimetallic
compounds are qualitatively viewed within the spin-wave treatment,
one of which successfully grows into a bulk magnet, while the other
of which ends in a singlet ground state.
Then, concentrating on the ferrimagnetic arrangement on a two-leg
ladder which is well indicative of general coupled-chain
ferrimagnets, we develop the spin-wave theory and fully reveal its
low-energy structure.
We inquire further into the ferromagnetic aspect of the ferrimagnetic
ladder numerically calculating the sublattice magnetization and the
magnetic susceptibility.
There exists a moderate coupling strength between the chains in order
to obtain the most ferromagnetic ferrimagnet.
\end{abstract}
\pacs{PACS numbers: 75.10.Jm, 75.50.Gg, 75.30.Ds, 75.40.Mg}
\narrowtext

\section{Introduction}\label{S:I}

   Much effort has been devoted to designing molecular systems
ordering ferromagnetically and recent progress \cite{OK095} in the
molecular chemistry in this context deserves special mention.
One unique approach \cite{JM769} to molecular ferromagnets consists
of assembling molecular bricks so as to construct a low-dimensional
system whose ground state possesses a nonzero resultant spin and then
coupling the chains or the layers in a ferromagnetic fashion.
We are fully convinced that ferromagnetic couplings between
nearest-neighbor magnetic centers lead to the highest spin
multiplicity, but such an interaction generally requires rather high
site symmetries realizing the orthogonality of the magnetic orbitals.
Except for a few polynuclear metal complexes
\cite{OK165,YJ585,OK557,AB695} with the symmetry-imposed
orthogonality, most of ferromagnetically coupled molecular systems
\cite{CC001} critically depend on some structural parameters which
are hard to handle chemically.
An alternative solution \cite{OK089} to a highly magnetic ground
state was obtained from a rather different
concept$-$antiferromagnetically coupled polymetallic systems with
irregular spin-state structures.
It is the noncompensation of local spins, rather than the
ferromagnetic interaction between nearest neighbors, that a magnetic
ground state really demands.
Ordered bimetallic chain compounds \cite{AG727} are thus synthesized
and since then the magnetic properties of ferrimagnetic chains have
extensively been investigated \cite
{MV144,MD992,SP707,SY610,SY024,SY211,SY795,TO576,TK762,TK813,NI024,CW057}.

   This new strategy for molecular ferromagnets is well described by
two contrastive but family compounds \cite{YP138,OK782,PK325}:
MnCu(pba)(H$_2$O)$_3$$\cdot$$2$H$_2$O with
 pba = \ 1,3-propylenebis(oxamato) \
 = \ C$_7$H$_6$N$_2$O$_6$ and
MnCu(pbaOH)(H$_2$O)$_3$ with
 pbaOH = 2-hydroxy-1,3-propylenebis(oxamato)
 = C$_7$H$_6$N$_2$O$_7$,
which are hereafter abbreviated as ``pba" and ``pbaOH" complexes,
respectively.
Both materials contain one-dimensional alignments of alternating
magnetic centers Mn of spin $S=\frac{5}{2}$ and Cu of spin
$s=\frac{1}{2}$ and behave as chain compounds of local spins $S-s$
coupled ferromagnetically.
However, due to the interchain interactions, their three-dimensional
behaviors are qualitatively different:
With decreasing temperature, the divergence of the magnetic
susceptibility, which is reminiscent of one-dimensional ferromagnets,
is stopped at the onset of three-dimensional ordering for the pba
complex, whereas the pbaOH complex still behaves as a ferromagnet
below the critical temperature.
The former ends up with a singlet ground state in three dimension,
while the latter really achieves a molecule-based ferromagnet.
Besides the bimetallic chain compounds, certain metal-radical mixed
materials \cite{AC756,AC976,AC940} have also achieved the latter
situation.

   In such circumstances, Heisenberg alternating-spin chains were
fully studied and their ferromagnetic-antiferromagnetic mixed
features \cite{SY033} were illuminated.
However, very little has been calculated beyond one dimension.
Namely, theoreticians have indeed assembled molecular bricks so as to
construct ferrimagnetic chains but have not yet coupled them in a
ferromagnetic fashion.
Thus, in this article, we proceed to the latter part of the crystal
engineering of molecular ferromagnets$-$interchain coupling.
First, taking the case of the $ab$ planes of the pba and pbaOH
complexes, we generally discuss the low-energy structures of
mixed-spin systems beyond one dimension.
Second, focusing on the ferromagnetic ordering of the ferrimagnetic
chains, we fully investigate two-leg mixed-spin ladders, which are
well indicative of general ferrimagnetic coupled-chain systems.
Calculating their thermal, as well as quantum, behavior, we finally
understand how effective the interchain coupling is in designing
molecular ferromagnets.
We make full use of the spin-wave theory employing numerical tools as
well.

\begin{figure}
\begin{flushleft}
\vspace*{-15mm}
\mbox{\qquad\psfig{figure=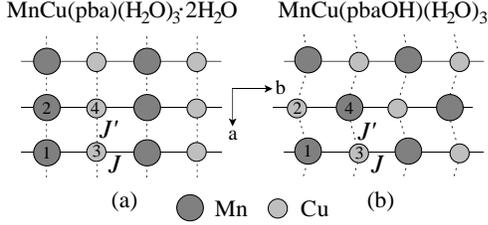,width=85mm,angle=0}}
\end{flushleft}
\caption{Schematic representation of the crystal structures of
         MnCu(pba)(H$_2$O)$_3$$\cdot$$2$H$_2$O (a) and
         MnCu(pbaOH)(H$_2$O)$_3$ (b) in their $ab$ planes.}
\label{F:structure}
\end{figure}

\section{Mixed-Spin Coupled-Chain Systems}\label{S:MSCCS}

   The above-mentioned bimetallic chain compounds,
MnCu(pba)(H$_2$O)$_3$$\cdot$$2$H$_2$O and
MnCu(pbaOH)(H$_2$O)$_3$,
are schematized in Fig. \ref{F:structure}.
Due to their characteristic crystal packings \cite{OK782}, the
dominant interchain interaction occurs between spins of the same kind
in pba, whereas between those of different kinds in pbaOH.
Thus these materials may essentially be described by the
$l$-leg-$L$-rung mixed-spin coupled-chain Hamiltonians
\begin{eqnarray}
   {\cal H}^{(\alpha)}=
   &&
   \sum_{i=1}^l\sum_{j=1}^L
   \Bigl(
    J \mbox{\boldmath$\sigma$}_{i,j  }^{(\alpha)}\cdot
      \mbox{\boldmath$\sigma$}_{i,j+1}^{(\alpha)}
   +J'\mbox{\boldmath$\sigma$}_{i  ,j}^{(\alpha)}\cdot
      \mbox{\boldmath$\sigma$}_{i+1,j}^{(\alpha)}
   \Bigr)\,,
   \label{E:H}
\end{eqnarray}
where a and b are set for the index $\alpha$, corresponding to pba
and pbaOH, respectively, and the $j$th magnetic center on the $i$th
chain, $\mbox{\boldmath$\sigma$}_{i,j}^{(\alpha)}$ is explicitly
given as
\begin{eqnarray}
   &&
   \mbox{\boldmath$\sigma$}_{2m  ,2n-1}^{\rm(a)}
  =\mbox{\boldmath$\sigma$}_{2m-1,2n-1}^{\rm(a)}
  =\mbox{\boldmath$S$}\,,\ \ 
   \mbox{\boldmath$\sigma$}_{2m  ,2n  }^{\rm(a)}
  =\mbox{\boldmath$\sigma$}_{2m-1,2n  }^{\rm(a)}
  =\mbox{\boldmath$s$}\,,
   \nonumber \\
   &&
   \mbox{\boldmath$\sigma$}_{2m-1,2n-1}^{\rm(b)}
  =\mbox{\boldmath$\sigma$}_{2m  ,2n  }^{\rm(b)}
  =\mbox{\boldmath$S$}\,,\ \ 
   \mbox{\boldmath$\sigma$}_{2m-1,2n  }^{\rm(b)}
  =\mbox{\boldmath$\sigma$}_{2m  ,2n-1}^{\rm(b)}
  =\mbox{\boldmath$s$}\,,
   \nonumber
\end{eqnarray}
with $\mbox{\boldmath$S$}$ and $\mbox{\boldmath$s$}$ representing
Mn(II) of spin $S=\frac{5}{2}$ and Cu(II) of spin $s=\frac{1}{2}$,
respectively.
We adopt the periodic boundary condition in both leg and rung
directions.

   In order to qualitatively investigate the low-energy structures,
we here make a spin-wave approach to the models.
Assuming the N\'eel configuration and then introducing bosonic
spin-deviation operators of four kinds, where the numbering notation
is specified in Fig. \ref{F:structure}, we obtain the spin-wave
Hamiltonians up to quadratic order in the momentum space as
\begin{eqnarray}
   &&
   {\cal H}_{\rm SW}^{\rm (a)}
    =J\sum_{\mbox{\boldmath$k$}}
     \Bigl[
       {\bf A}_{\mbox{\boldmath$k$}}^\dagger
       {\cal H}_{\mbox{\boldmath$k$}}^{\rm (a)}
       {\bf A}_{\mbox{\boldmath$k$}}
      -4Ss-2r(S^2+s^2)
     \Bigr]\,,
   \label{E:SWHa}
   \\
   &&
   {\cal H}_{\rm SW}^{\rm (b)}
    =J\sum_{\mbox{\boldmath$k$}}
     \Bigl[
       {\bf A}_{\mbox{\boldmath$k$}}^\dagger
       {\cal H}_{\mbox{\boldmath$k$}}^{\rm (b)}
       {\bf A}_{\mbox{\boldmath$k$}}
      -4(1+r)Ss
     \Bigr]\,,
   \label{E:SWHb}
\end{eqnarray}
where
\begin{eqnarray}
   &&
   {\bf A}_{\mbox{\boldmath$k$}}
    ={\rm T}
     \left[
      a_{\mbox{\boldmath$k$}}^{(1)}\,\cdots\,
      a_{\mbox{\boldmath$k$}}^{(4)}\,
      a_{\mbox{\boldmath$k$}}^{(1)\dagger}\,\cdots\,
      a_{\mbox{\boldmath$k$}}^{(4)\dagger}
     \right]\,,
   \\
   &&
   {\cal H}_{\mbox{\boldmath$k$}}^{(\alpha)}
    =\left[
     \begin{array}{cc}
      {\Gamma}^{(\alpha)} &
      {\Delta}_{\mbox{\boldmath$k$}}^{(\alpha)} \\
      {\Delta}_{\mbox{\boldmath$k$}}^{(\alpha)} &
      {\Gamma}^{(\alpha)}
     \end{array}
     \right]\,,
\end{eqnarray}
with
\widetext
\begin{eqnarray}
   &&
   {\Gamma}^{\rm (a)}
    ={\rm diag}
     \big[
      s+rS, s+rS, S+rs, S+rs
     \big]\,,\ \ 
   {\Delta}_{\mbox{\boldmath$k$}}^{\rm(a)}
    =\left[
      \begin{array}{cccc}
       0 & rS{\rm cos}\frac{k_y}{2} &
       \sqrt{Ss}{\rm cos}\frac{k_x}{2} & 0 \\
       rS{\rm cos}\frac{k_y}{2} & 0 &
       0 & \sqrt{Ss}{\rm cos}\frac{k_x}{2} \\
       \sqrt{Ss}{\rm cos}\frac{k_x}{2} & 0 &
       0 & rs{\rm cos}\frac{k_y}{2} \\
       0 & \sqrt{Ss}{\rm cos}\frac{k_x}{2} &
       rs{\rm cos}\frac{k_y}{2} & 0
      \end{array}
     \right]\,,
   \\
   &&
   {\Gamma}^{\rm (b)}
    =(1+r){\rm diag}
     \big[
      s, S, S, s
     \big]\,,\ \ 
   {\Delta}_{\mbox{\boldmath$k$}}^{\rm(b)}
    =\sqrt{Ss}
     \left[
      \begin{array}{cccc}
       0 & r{\rm cos}\frac{k_y}{2} &
       {\rm cos}\frac{k_x}{2} & 0 \\
       r{\rm cos}\frac{k_y}{2} & 0 &
       0 & {\rm cos}\frac{k_x}{2} \\
       {\rm cos}\frac{k_x}{2} & 0 &
       0 & r{\rm cos}\frac{k_y}{2} \\
       0 & {\rm cos}\frac{k_x}{2} &
       r{\rm cos}\frac{k_y}{2} & 0
      \end{array}
     \right]\,.
\end{eqnarray}
\narrowtext
Here, we have set twice the lattice constants in the $x$ (leg) and
$y$ (rung) directions both equal to unity, whereas $J'/J$ equal to
$r$.
The spin-wave excitations are revealed through the Bogoliubov
transformation \cite{TF398}, which can be described in terms of a
$8\times 8$ matrix ${\cal U}$ as
\begin{equation}
   {\bf A}_{\mbox{\boldmath$k$}}
   ={\cal U}{\bf B}_{\mbox{\boldmath$k$}}\,;\ \ 
   {\cal U}{\cal N}{\cal U}^\dagger
   ={\cal U}^\dagger{\cal N}{\cal U}
   ={\cal N}\,,
   \label{E:BT}
\end{equation}
where the metric
\begin{equation}
   {\cal N}={\rm diag}[1,1,1,1,-1,-1,-1,-1]\,,
\end{equation}
has been introduced in order to demand that
\begin{equation}
   {\bf B}_{\mbox{\boldmath$k$}}
    ={\rm T}
     \left[
      b_{\mbox{\boldmath$k$}}^{(1)}\,\cdots\,
      b_{\mbox{\boldmath$k$}}^{(4)}\,
      b_{\mbox{\boldmath$k$}}^{(1)\dagger}\,\cdots\,
      b_{\mbox{\boldmath$k$}}^{(4)\dagger}
     \right]\,,
\end{equation}
should satisfy the boson commutation relations.

   In the case (a) we obtain two spin-wave modes, which are both
doubly degenerate, as
\begin{equation}
   \omega_{\sigma}^{\rm(a)}(\mbox{\boldmath$k$})
   =\sqrt{2(A+\sigma\sqrt{B})}\,,
   \label{E:omegaa}
\end{equation}
where the index $\sigma$ takes $\pm$ and
\begin{eqnarray}
   &&
   \frac{A}{J^2}=(S^2+s^2)
     \Bigl(1+r^2{\rm sin}^2\frac{k_y}{2}\Bigr)
    +4Ss
     \Bigl(
      r-\frac{1}{2}{\rm cos}^2\frac{k_x}{2}
     \Bigr)\,,
   \nonumber \\
   &&
   \frac{B}{J^4}=(S^2-s^2)^2
     \Bigl(
      1-r^2{\rm sin}^2\frac{k_y}{2}
     \Bigr)^2
   \nonumber \\
   &&\quad
    -4Ss{\rm cos}^2\frac{k_x}{2}
     \Bigl[
      (1-r)^2(S-s)^2
     -r^2(S+s)^2{\rm cos}^2\frac{k_y}{2}
     \Bigr]\,,
   \nonumber \\
   &&
\end{eqnarray}
\widetext
\begin{figure}
\vspace*{-10mm}
\qquad\qquad\mbox{\psfig{figure=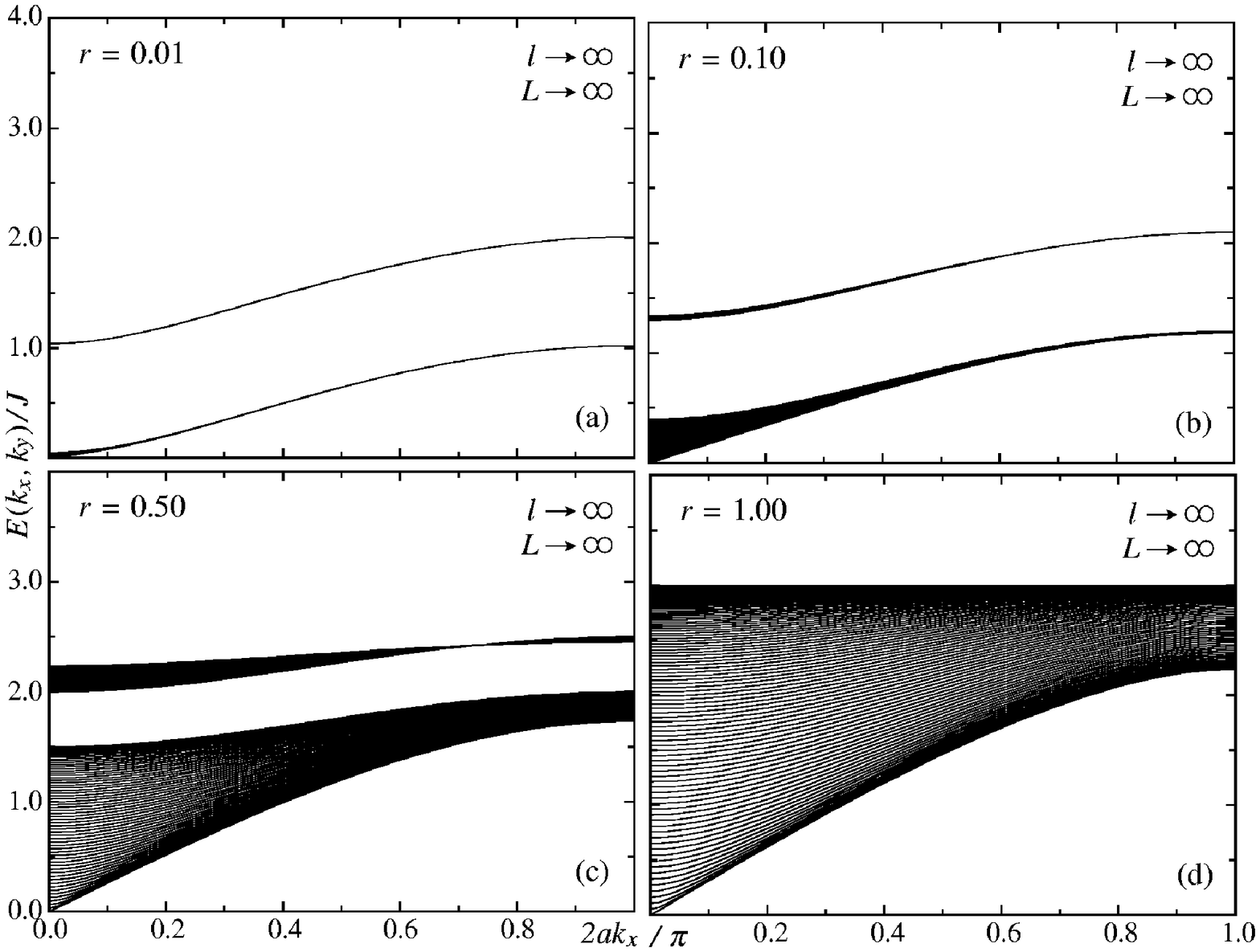,width=140mm,angle=0}}
\caption{Dispersion relations of the spin-wave excitations as
         functions of $k_x$ for the Hamiltonian ${\cal H}^{\rm(a)}$
         with $k_y$ fully running in the Brillouin zone and $a$
         denoting the lattice constant in the leg direction.
         $(S,s)$ is set equal to $(1,\frac{1}{2})$.
         $r$ is increased in the order (a) to (d).}
\label{F:dsp2Ds}
\end{figure}
\begin{figure}
\vspace*{-5mm}
\qquad\qquad\mbox{\psfig{figure=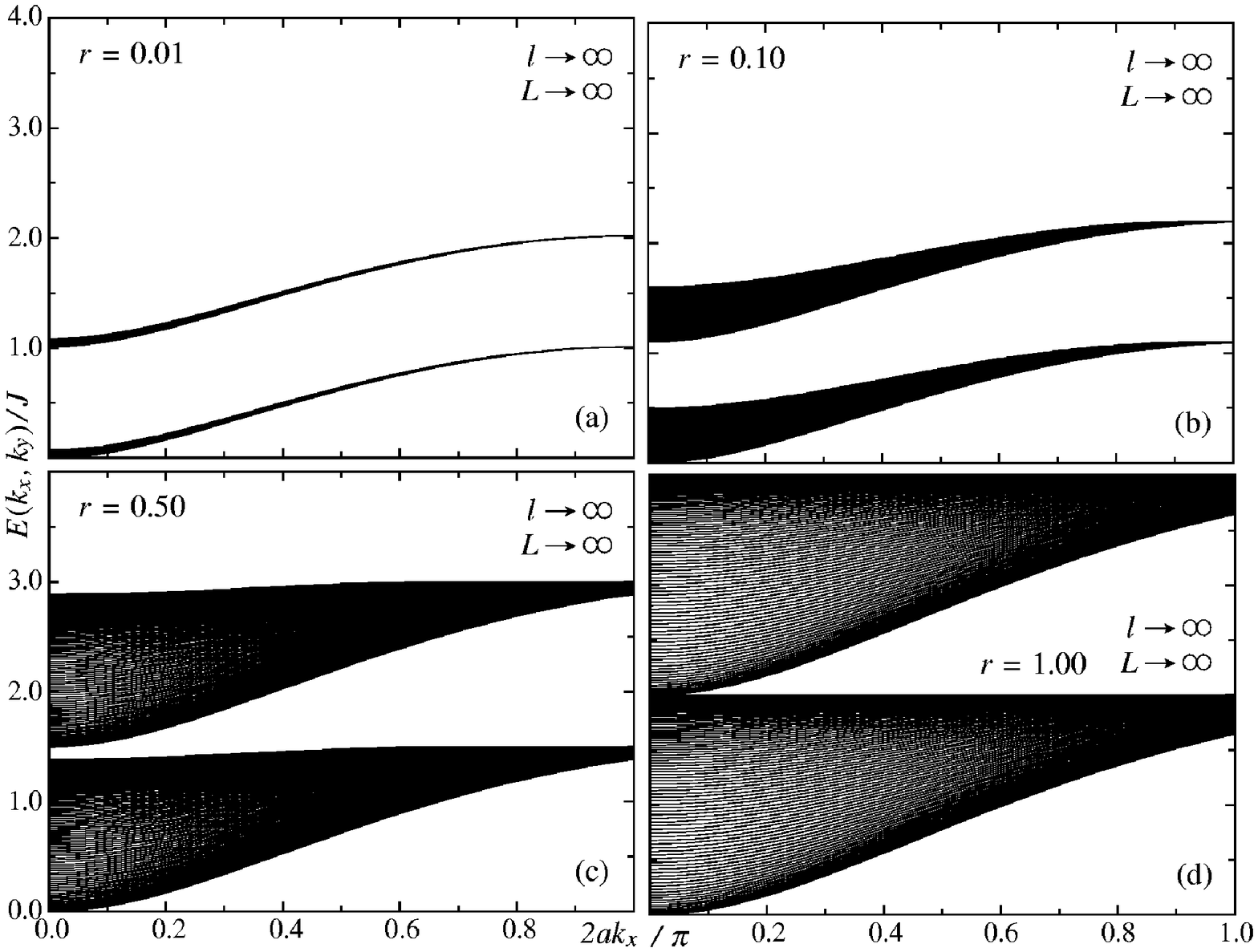,width=140mm,angle=0}}
\caption{Dispersion relations of the spin-wave excitations as
         functions of $k_x$ for the Hamiltonian ${\cal H}^{\rm(b)}$
         with $k_y$ fully running in the Brillouin zone and $a$
         denoting the lattice constant in the leg direction.
         $(S,s)$ is set equal to $(1,\frac{1}{2})$.
         $r$ is increased in the order (a) to (d).}
\label{F:dsp2D}
\end{figure}
\narrowtext
while in the case (b) the spin waves are generally composed of four
modes as
\begin{equation}
   \omega_{\sigma\tau}^{\rm(b)}(\mbox{\boldmath$k$})
   =\sqrt{C-D_\tau}+\sigma\sqrt{C}\,,
   \label{E:omegab}
\end{equation}
where the indices $\sigma$ and $\tau$ take $\pm$ and
\begin{eqnarray}
   &&
   \frac{C}{J^2}=(1+r)^2(S+s)^2\,,
   \nonumber \\
   &&
   \frac{D_\tau}{J^2}=4Ss
     \Bigl(
      {\rm cos}\frac{k_x}{2}-\tau r{\rm cos}\frac{k_y}{2}
     \Bigr)^2\,.
\end{eqnarray}
In Figs. \ref{F:dsp2Ds} and \ref{F:dsp2D} we schematically show the
dispersion relations as functions of $k_x$ running $k_y$ from $0$ to
$2\pi$.
Since quasi-one-dimensional situations are of our interest, here and
in the following we explicitly observe the dispersion in the leg
direction.
Although the excitation spectra (\ref{E:omegaa}) and (\ref{E:omegab})
look alike in the small-$r$ region, there occurs an essential
difference between them with the inclusion of the interchain
couplings.
In the case (a) the lowest excitation mode exhibits a linear
dispersion for small momenta as
\begin{equation}
   \omega_{-}^{\rm(a)}(k_x,k_y=0)\sim v_x^{\rm(a)}k_x\,,
\end{equation}
with
\begin{equation}
   v_x^{\rm(a)}
    =J\sqrt{2rSs\frac{(S-s)^2+2(1+r)Ss}{(S-s)^2+4rSs}}\,,
\end{equation}
whereas in the case (b) it remains quadratic as
\begin{equation}
   \omega_{--}^{\rm(b)}(k_x,k_y=0)\sim v_x^{\rm(b)}k_x^2\,,
\end{equation}
with
\begin{equation}
   v_x^{\rm(b)}=\frac{SsJ}{2(S-s)}\,.
\end{equation}
The velocity $v_x^{\rm(a)}$ monotonically increases with $r$, while
the curvature $v_x^{\rm(b)}$ does not depend on $r$ up to $O(S^1)$.
The low-energy physics is much more sensitive to the interchain
interaction of the type (a).

   Thus, we find that the three-dimensional ground state of the pba
complex is singlet regardless of the magnitude of $r$, while the
pbaOH complex can be a three-dimensional magnet.
Systems of the type (a) miss our goal$-$molecular ferromagnets, but
they are interesting in relation to the Haldane-gap problem
\cite{FH464,FH153}.
Coexistence of Haldane-gap excitations and gapless ones induced by
the long-range order in $R_2$BaNiO$_5$
($R=\mbox{magnetic\ rare\ earth}$)
\cite{AZ437,AZ210,TY516,TY424,SR382} has been motivating theoretical
approaches to the model ${\cal H}^{\rm(a)}$ from the
$r\rightarrow\infty$ limit \cite{SM068,SM786,AK133,YT189}.
Now we restrict ourselves to the model ${\cal H}^{\rm(b)}$ and
explore further into the low-energy physics essential to
ferrimagnetic coupled-chain systems.
\widetext
\begin{figure}
\qquad\qquad\mbox{\psfig{figure=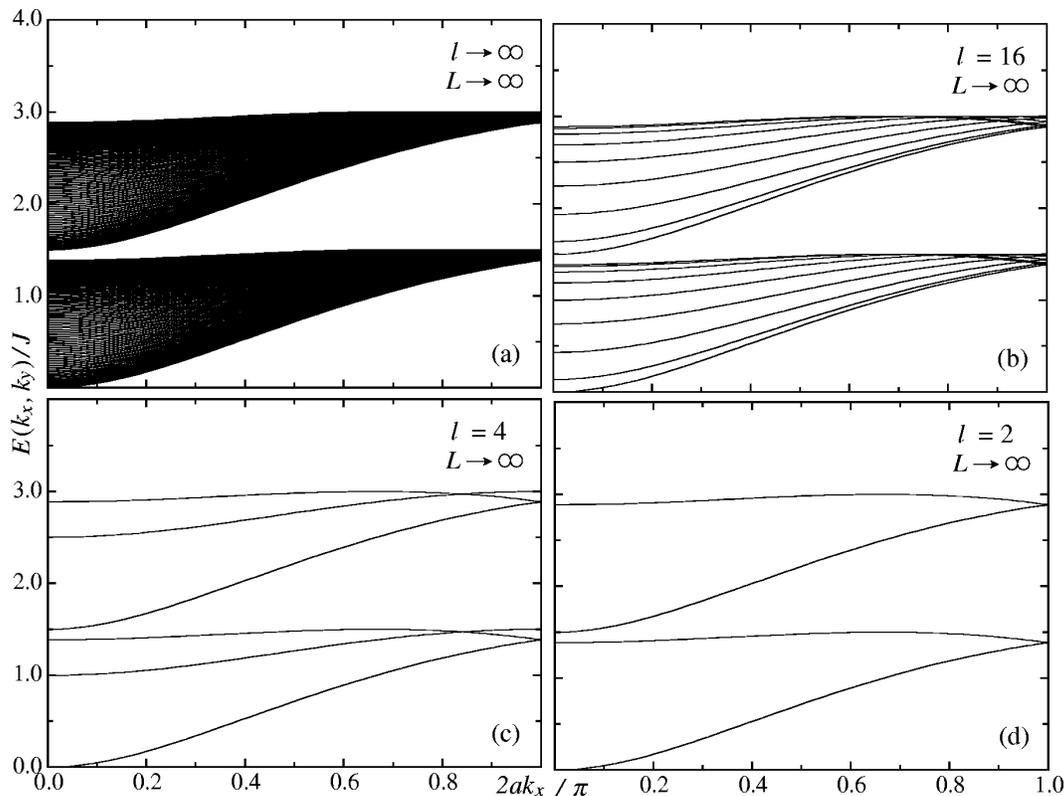,width=140mm,angle=0}}
\caption{Dispersion relations of the spin-wave excitations as
         functions of $k_x$ for the Hamiltonian ${\cal H}^{\rm(b)}$
         with $k_y$ taking all the possible values and $a$ denoting
         the lattice constant in the leg direction.
         $(S,s)$ is set equal to $(1,\frac{1}{2})$, while $r$ equal
         to $0.5$.
         The number of legs, $l$, is decreased in the order (a) to
         (d).}
\label{F:dsplleg}
\end{figure}
\begin{figure}
\vspace*{-5mm}
\qquad\qquad\mbox{\psfig{figure=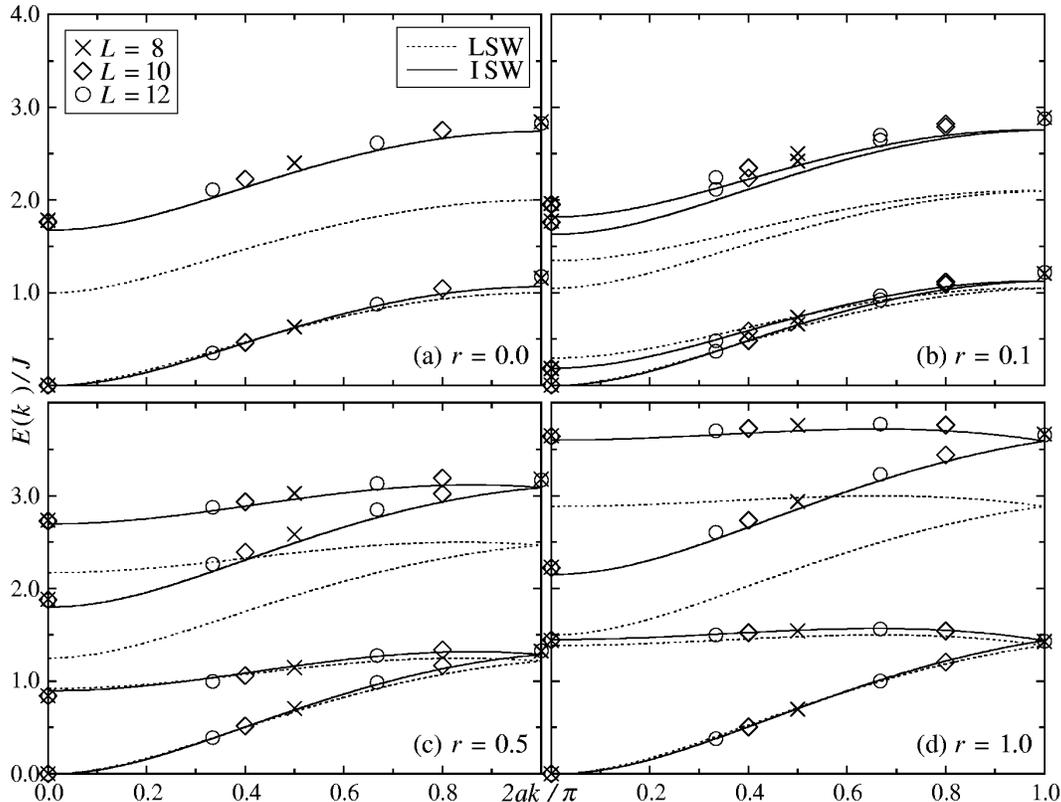,width=140mm,angle=0}}
\caption{Dispersion relations of the ferromagnetic and
         antiferromagnetic elementary excitations, which reduce and
         enhance the ground-state magnetization and thus lie in the
         subspaces of magnetization $M=(S-s)L\mp 1$, respectively,
         for the $(S,s)=(1,\frac{1}{2})$ two-leg ferrimagnetic
         ladders.
         The noninteracting-spin-wave (LSW) and interacting-spin-wave
         (ISW) calculations are shown by dotted and solid lines,
         respectively, whereas symbols represent the
         exact-diagonalization results.
         $r$ is increased in the order (a) to (d).}
\label{F:dsp2leg}
\end{figure}
\narrowtext

\section{Ferrimagnetic Spin Ladders}\label{S:FSL}

   Considering the present aim of revealing essential effects of
the interchain interaction on quasi-one-dimensional ferrimagnetic
phenomena, further calculations of the two-dimensional lattice are
not necessarily efficient.
Though numerical tools are indispensable for quantitative
understanding, the large degrees of freedom strongly reduce their
feasibility.
In this context, Fig. \ref{F:dsplleg} gives us a useful piece of
information.
It is true that the number of states decreases with decreasing $l$,
but the essential band structure remains unchanged.
Even in the two-leg ladder system,
the band width,
$\omega_{\sigma +}^{\rm(b)}(k_x=2{\rm cos}^{-1}r,k_y=0)
-\omega_{\sigma -}^{\rm(b)}(k_x=k_y=0)$,
the curvature of the ferromagnetic dispersion, $v_x^{\rm(b)}$, and
the gap of the lowest antiferromagnetic excitation mode,
$\omega_{+-}^{\rm(b)}(0)$,
they are all exactly the same as those of the corresponding two- and
three-dimensional systems within the up-to-$O(S^1)$ spin-wave theory.
The two-leg ferrimagnetic ladder is thus found to be very well
indicative of any ferrimagnet in higher dimensions.
From the physical point of view, ladder ferrimagnets are highly
interesting in themselves, where quantum effects are much more
remarkable and therefore novel phenomena are much more expected.
There are in fact recent reports \cite{AL343,AL725} on quantum
ferrimagnetic spin ladders where quantized plateaux in their
ground-state magnetization curves have been revealed.

   We proceed to detailed calculation of the low-energy structure
of the ferrimagnetic spin ladder.
Due to the periodic boundary condition imposed on Eq. (\ref{E:H}),
the two-leg ladder with an intrachain coupling $J$ and an interchain
coupling $J'$ is described by the Hamiltonian ${\cal H}^{\rm(b)}$
with $l=2$ and $J'$ being replaced by $J'/2$.
For the later convenience, we take $N$ for the number of the elementary
units, $L/2$, and again adopt the notation $r=J'/J$.
The dispersion relations of the free spin waves are compared with the
exact-diagonalization results in Fig. \ref{F:dsp2leg}.
The lower-bands of ferromagnetic features are very well described by
the linear spin-wave theory, whereas the upper-bands of
antiferromagnetic features significantly deviate from the
free-spin-wave excitations.
The relatively poor description of the antiferromagnetic branches by
the free spin waves implies that the quantum effect is more relevant
in the antiferromagnetic excitations.
This is rather convincing when we consider the conventional spin-wave
description of Hisenberg ferromagnets and antiferromagnets.
The ferromagnetic spin waves correctly describe the low-lying
excitations, whereas the antiferromagnetic ones merely give a
qualitative view of the low-energy structure leading to the
divergence of sublattice magnetizations.
Here, the spin-wave theory is much more potential and the spin-wave
series can in principle reach an accurate description of the
antiferromagnetic, as well as ferromagnetic, excitations owing to
nondivergent sublattice magnetizations.

   In order to refine the up-to-$O(S^1)$ dispersion relations
(\ref{E:omegab}), let us consider interactions between the spin
waves.
We may rediagonalize the one-body Hamiltonian (\ref{E:SWHb}) together
with the two-body terms of the order $O(S^0)$ in the naivest attempt
to go beyond the linear spin-wave theory.
However, such a treatment so misreads the low-energy structure as to
bring a gap to the lowest-lying ferromagnetic excitation branch.
Therefore, preserving the Bogoliubov transformation, we pick up
relevant contributions to the dispersions, as well as to the
ground-state energy, from the $O(S^0)$ terms.
The Wick theorem allows us to rewrite the spin-wave Hamiltonian of
the present two-leg ladder system as
\begin{eqnarray}
   &&
   {\cal H}_{\rm SW}
    =E_{\rm class}+E_0+E_1
    +{\cal H}_{\rm irrel}+{\cal H}_{\rm quart}+O(S^{-1})
   \nonumber \\
   &&\quad
    +\sum_{\sigma,\tau=\pm}\sum_k
     \big[
      \omega_{\sigma\tau}(k)+\delta\omega_{\sigma\tau}(k)
     \big]
     b_k^{(i(\sigma,\tau))\dagger}b_k^{(i(\sigma,\tau))}\,,
\end{eqnarray}
where the index $i$ of the bosonic operators is a function of
$\sigma$ and $\tau$ as $i(\sigma,\tau)=2(\sigma+1)/2+(\tau+1)/2+1$,
taking $1$ to $4$.
${\cal H}_{\rm irrel}$ contains irrelevant terms such
as $b_k^{(i)}b_k^{(j)}$ and ${\cal H}_{\rm quart}$ is the residual
two-body interactions in the normal order, both of which are
neglected so as to keep the lowest ferromagnetic excitation gapless.
$E_{\rm class}=-4NJ(1+r/2)Ss$ is the N\'eel-state energy, while $E_0$
and $E_1$ are the $O(S^1)$ and $O(S^0)$ quantum corrections to it,
respectively.
$\omega_{\sigma\tau}(k)$'s are the dispersions for the free spin
waves, whereas $\delta\omega_{\sigma\tau}(k)$'s are the $O(S^0)$
corrections to them.
They are all explicitly given in Appendix \ref{A:SWH}.
Thus-obtained up-to-$O(S^0)$ dispersion relations potentially gives a
precise description of the low-temperature thermodynamics
\cite{SY033}.

   The interacting-spin-wave dispersions are also shown in Fig.
\ref{F:dsp2leg}.
They excellently describe the antiferromagnetic, as well as
ferromagnetic, excitations.
We emphasize that the present highly accurate spin-wave description
of the low-energy structure is stably obtained for ferrimagnetic
chains, ladders, and layers with an arbitrary combination of $S$ and
$s$.
For $S=s$, where the present system is no more a ferrimagnet, the
quantum corrections (\ref{E:EgS0}) and (\ref{E:omegaS0}) diverge.
This is due to the divergence of the quantum spin reduction
\begin{equation}
   \tau=\frac{1}{N}\sum_k
        \langle
         a_k^{(i)\dagger}a_k^{(i)}
        \rangle_{\rm g}\,,
   \label{E:taudef}
\end{equation}
where $\langle\ \rangle_{\rm g}$ means the ground-state average.
The conventional spin-wave theory \cite{PA694,RK568} for
low-dimensional Heisenberg antiferromagnets is plagued by the
difficulty of the zero-field sublattice magnetizations diverging.
Ferrimagnetic systems are generally free from this difficulty and
therefore quantum correlations can systematically be calculated.
Table \ref{T:Eg} further demonstrates the excellence of the
ferrimagnetic spin-wave theory.
Unless $r$ is too large to maintain the quasi-one-dimensional aspect,
the interchain coupling contributes to the enhancement of the
effective dimension of the system.
With increasing $r$, the N\'eel configuration indeed better
approximates the ground state.
Although the classical configuration considerably deviates from the
exact ground state at the decoupled-chain limit $r=0$, the spin-wave
series still well describe the ground-state correlations.
\section{Ferromagnetic Features of Ferrimagnets}\label{S:FFF}

\begin{figure}
\begin{flushleft}
\ \mbox{\psfig{figure=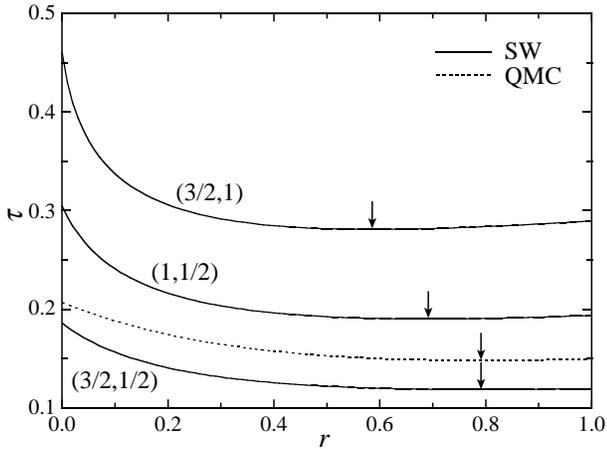,width=82mm,angle=0}}
\end{flushleft}
\vskip -67mm
\caption{Spin-wave calculations of the quantum spin reduction $\tau$
         as a function of $r$ for the two-leg ferrimagnetic ladders
         with various values of $(S,s)$.
         Quantum Monte Carlo estimates are also shown at
         $(S,s)=(1,\frac{1}{2})$ for reference.
         The minima are indicated by arrows.}
\label{F:tau}
\end{figure}

   How should we couple the ferrimagnetic chains in order to grow
their ferromagnetic features?
We further consider the interchain-coupling effect from this point of
view.
The ground-state magnetization must be a measure for magnets and it
is therefore interesting to observe the quantum spin reduction $\tau$
as a function of $r$.
Equation (\ref{E:taudef}) is calculated as
\begin{equation}
   \tau=\frac{1}{N}\sum_k
        \big(
         \psi_1^{(2)}(k)^2+\psi_2^{(2)}(k)^2
        \big)\,,
   \label{E:taucal}
\end{equation}
within the present spin-wave treatment and is plotted in Fig.
\ref{F:tau} in comparison with the quantum Monte Carlo estimates.
When we consider $\tau$ as a function of constituent spins, Eq.
(\ref{E:taucal}) monotonically decreases as $S/s$ increases,
diverging at the antiferromagnetic limit $S/s=1$ and vanishing for
the ferromagnetic limit $S/s\rightarrow\infty$.
Though the divergence at $S=s$ is totally due to the
Holstein-Primakoff transformation, this observation of $\tau$ fully
justifies the view that $\tau$ should be a measure for the
ferromagnetic aspect of ferrimagnets.

   The spin waves overestimate quantum fluctuations but they
successfully reproduce the minimum of $\tau$ as a function of $r$.
When we assemble molecular bricks so as to construct ferrimagnetic
chains and then couple them in a ferromagnetic fashion, the N\'eel
configuration grows with increasing $r$ but there is a {\it moderate}
point $r_{\rm c}$ to maximize that.
For an arbitrary $l$-leg ladder ferrimagnet,
$\tau(r\rightarrow\infty)$ is generally larger than $\tau(r=0)$ and
thus there should be a minimum of $\tau$ as a function of $r$.
The minimum point $r_{\rm c}$ is smaller than unity as far as $l$
stays finite.
If we consider the simple square-lattice plane as is illustrated in
Fig. \ref{F:structure}, $r_{\rm c}$ approaches unity for
$l\rightarrow\infty$.
On the other hand, the recent progress \cite{YH924} in the molecular
chemistry may raise even a possibility of mixed-spin chains based on
stable organic radicals forming into a honeycomb lattice, where
$r_{\rm c}$ remains smaller than unity even for $l\rightarrow\infty$.
Furthermore the synthesis of a ladder ferrimagnet itself is now
indeed in progress \cite{YH}.
$r_{\rm c}$ is sensitive to the constituent spins as well as the
crystal structure.
This point should also be taken into consideration in designing
molecular ferromagnets.
$r_{\rm c}$ is an increasing function of the ratio $g\equiv S/s$,
suggesting that ferrimagnets should be characterized by $g$ rather
than $S$ and $s$ themselves.
As for the bimetallic chain compounds
MM$'$(pbaOH)(H$_2$O)$_3$$\cdot$$n$H$_2$O,
the larger-spin magnetic centers were systematically tuned from
Mn ($S=\frac{5}{2}$) to Ni ($S=1$), while the smaller-spin ones were
mainly fixed to Cu ($s=\frac{1}{2}$) \cite{PK325}.
This is partly due to the problem of crystal engineering but might
on the other hand aim at suppressing the quantum spin reduction
$\tau$.
However, as far as we work along this line, a rather strong
interchain coupling is needed to attain the highest transition
temperature.
If we replace both magnetic centers M and M$'$ simultaneously,
$r_{\rm c}$ can relatively be reduced at the cost of the amplitude of
$\tau$ and therefore the highest transition temperature might be
obtained at a smaller feasible interchain coupling.
\vskip 4mm
\vskip 4mm
\begin{figure}
\begin{flushleft}
$\,$\mbox{\psfig{figure=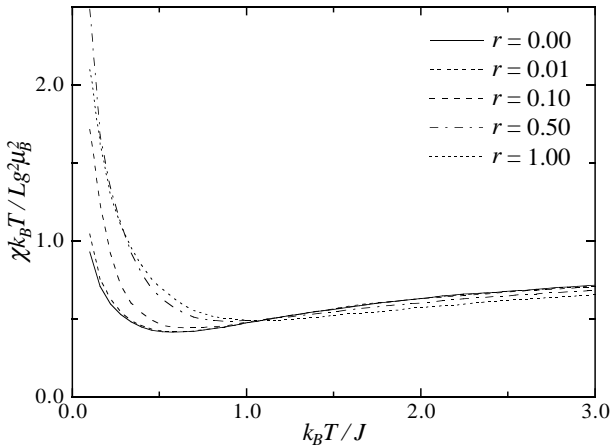,width=83mm,angle=0}}
\end{flushleft}
\vskip 0mm
\caption{Quantum Monte Carlo calculations of the temperature
         dependence of the magnetic susceptibility times temperature
         for the $(S,s)=(1,\frac{1}{2})$ two-leg ferrimagnetic
         ladders.}
\label{F:chiT}
\end{figure}
\vskip 2mm

   In relation to $\tau$, let us observe the ferrimagnetic
susceptibility in Fig. \ref{F:chiT}.
The susceptibility-temperature product $\chi T$ is a monotonically
decreasing function of $T$ in ferromagnets, while a monotonically
increasing function  in antiferromagnets.
Therefore, the present observation may be regarded as a
ferromagnetic-antiferromagnetic mixed nature of ferrimagnets.
The low-temperature divergence is really reminiscent  of the
ferromagnetic susceptibility and therefore signifies {\it how
ferromagnetic the ferrimagnet is}.
Interestingly enough, as $r$ moves away from $0$ toward $1$, the
diverging behavior is indeed sharpened in the beginning but reaches
its limit at a certain value of $r$.
The divergence at $r=1$ is clearly duller than that at $r=0.5$.
Though the diverging susceptibility is very hard to calculate
accurately at low temperatures, {\it the most ferromagnetic
ferrimagnet} seems to be realized around $r=r_{\rm c}$.

\section{Summary and Discussion}\label{S:SD}

   Motivated by the ferrimagnetic chain compounds, we have
investigated the magnetic properties of mixed-spin coupled-chain
systems featuring a spin-wave analysis.
The present spin-wave description is available for general
ferrimagnets.
It is not only highly successful for quantum correlations but also
open to thermal properties complementing numerical investigations.
While the small-momentum ferromagnetic excitations should dominate
the low-temperature thermodynamics, exact-diagonalization methods are
totally unable to describe them.
Even with a quantum Monte Carlo technique \cite{SY609}, it is rather
hard to obtain the curvature $v$ of their quadratic dispersion
relation.
The present procedure readily gives $v$ as
\begin{eqnarray}
   &&
   \frac{v}{J}
    = \frac{Ss}{2(S-s)}
    - \frac{2(S+s)}{N(S-s)}
      \sum_k
      \big(
       \psi_1^{(2)}(k)^2+\psi_2^{(2)}(k)^2
      \big)
   \nonumber \\
   &&
    + \frac{2\sqrt{Ss}}{N(S-s)}
      \sum_k
      \big(
       \psi_1^{(2)}(k)\psi_1^{(4)}(k)
      +\psi_2^{(2)}(k)\psi_2^{(4)}(k)
      \big){\rm cos}\frac{k}{2}\,.
   \nonumber \\
   &&
\end{eqnarray}
It is free from $r$ within the linear spin-wave theory.
Its leading $r$ dependence arises from the $O(S^0)$ correction.
When $v$ is revealed, we can for instance obtain the leading behavior
of the specific heat at low temperatures as
\begin{equation}
   \frac{C}{Lk_{\rm B}}
    =\frac{3}{16}\zeta\Bigl(\frac{3}{2}\Bigr)
     \sqrt{\frac{k_{\rm B}T}{\pi v}}
    +O\Bigl(\frac{k_{\rm B}T}{J}\Bigr)\,,
\end{equation}
with Riemann's zeta function $\zeta(z)$, provided $r>k_{\rm B}T/J$.
The following terms of the order $O(k_{\rm B}T/J)$ can not
successfully be obtained within the present spin-wave formulation,
but a modified spin-wave theory \cite{SY008} will bring us them as
well as a precise description of the susceptibility.

   The present  calculation is not only a theoretical harvest but
also encourages further chemical explorations into molecular
ferromagnets.
While most of thus-far-synthesized ferrimagnets are inorganic-organic
hybrid materials, there may be genuine organic ferrimagnets, where
the variety and flexibility of the crystal structure, as well as
sufficiently small magnetic anisotropy, may enable us to investigate
quantum mixed-spin phenomena extensively and essentially.
An example approach \cite{DS342,MN503} consists of synthesizing an
alternating chain of a monoradical and a biradical with a triplet
ground state.
However, the low-temperature properties of the obtained compounds
deviate from the ferrimagnetic behavior suggesting the formation of
local singlet clusters.
In order to overcome such an inherent difficulty in organic
molecule-based materials, another strategy \cite{YH} is in progress,
where the crystal consists of a triradical containing both spin-$1$
and spin-$\frac{1}{2}$ units within itself.
This material is now believed to possess a ladder-like structure
rather than a chain structure predicted in the beginning of the
structural analysis.
Thus, the first example of a mixed-spin ladder has appeared, though
the ferrimagnetic ground state is still not realized there possibly
due to next nearest-neighbor interactions.

   Interchain interactions inevitably exist in low-dimensional
mixed-spin magnets and they seriously affect the formation of the
ferrimagnetic ground state.
Ladder systems are the simplest but efficient stage to study such
phenomena.
We expect a close collaboration between theoretical and experimental
investigations in designing molecule-based ferromagnets.

\acknowledgments

   The authors are grateful to Prof. T. Fukui for his useful comments
and fruitful discussion.
This work was supported by the Japanese Ministry of Education,
Science, and Culture through Grant-in-Aid No. 11740206 and by the
Sanyo-Broadcasting Foundation for Science and Culture.
The numerical computation was done in part using the facility of the
Supercomputer Center, Institute for Solid State Physics, University
of Tokyo.

\widetext
\begin{appendix}
\section{Spin-Wave Treatment of the Ferrimagnetic Ladder}
\label{A:SWH}

   We express the spin-wave Hamiltonian up to the order $O(S^0)$ as
\begin{equation}
   {\cal H}_{\rm SW}
   =E_{\rm class}+E_0+E_1
   +\sum_{\sigma,\tau=\pm}\sum_k
    \big[
     \omega_{\sigma\tau}(k)+\delta\omega_{\sigma\tau}(k)
    \big]
    b_k^{(i(\sigma,\tau))\dagger}b_k^{(i(\sigma,\tau))}\,,
\end{equation}
where
\begin{eqnarray}
   \frac{E_0}{J}
   =\sum_k
    &\biggl\{&
     \sqrt{\Bigl(1+\frac{r}{2}\Bigr)^2(S+s)^2
          -4Ss\big({\rm cos}\frac{k}{2}-\frac{r}{2}\big)^2}
    -\Bigl(1+\frac{r}{2}\Bigr)(S+s)
   \nonumber \\
    &+&
     \sqrt{\Bigl(1+\frac{r}{2}\Bigr)^2(S+s)^2
          -4Ss\Bigl({\rm cos}\frac{k}{2}+\frac{r}{2}\Bigr)^2}
    -\Bigl(1+\frac{r}{2}\Bigr)(S+s)
     \biggr\}\,,
   \label{E:EgS1}
\end{eqnarray}
\begin{eqnarray}
   \frac{E_1}{J}
   &=&
    - \frac{4}{N}
         \Bigl\{
          \sum_k
          \big(\psi_1^{(2)}(k)^2+\psi_2^{(2)}(k)^2\big)
         \Bigr\}^2
    - \frac{4}{N}
         \Bigl\{
          \sum_k
          \big(\psi_1^{(2)}(k)\psi_1^{(4)}(k)
              +\psi_2^{(2)}(k)\psi_2^{(4)}(k)\big)
          {\rm cos}\frac{k}{2}
         \Bigr\}^2
   \nonumber \\
   &-&
      \frac{2r}{N}
        \Bigl\{
         \sum_k
         \big(\psi_1^{(2)}(k)^2+\psi_2^{(2)}(k)^2\big)
        \Bigr\}^2
    - \frac{2r}{N}
        \Bigl\{
         \sum_k
         \big(\psi_1^{(1)}(k)\psi_1^{(2)}(k)
             +\psi_2^{(1)}(k)\psi_2^{(2)}(k)\big)
        \Bigr\}^2
   \nonumber \\
   &+& \frac{4}{N}
         \biggl(\sqrt{\frac{S}{s}}+\sqrt{\frac{s}{S}}\biggr)
          \sum_k
          \big(\psi_1^{(2)}(k)^2+\psi_2^{(2)}(k)^2\big)
          \sum_{k'}
          \big(\psi_1^{(2)}(k')\psi_1^{(4)}(k')
              +\psi_2^{(2)}(k')\psi_2^{(4)}(k')\big)
          {\rm cos}\frac{k'}{2}
   \nonumber \\
   &+& \frac{2r}{N}
         \biggl(\sqrt{\frac{S}{s}}+\sqrt{\frac{s}{S}}\biggr)
          \sum_k
          \big(\psi_1^{(2)}(k)^2+\psi_2^{(2)}(k)^2\big)
          \sum_{k'}
          \big(\psi_1^{(1)}(k')\psi_1^{(2)}(k')
              +\psi_2^{(1)}(k')\psi_2^{(2)}(k')\big)\,,
   \label{E:EgS0}
\end{eqnarray}
\begin{equation}
   \frac{\omega_{\sigma\tau}(k)}{J}
    =\sqrt{\Bigl(1+\frac{r}{2}\Bigr)^2(S+s)^2
          -4Ss\Bigl({\rm cos}\frac{k}{2}-\tau\frac{r}{2}\Bigr)^2}
    +\sigma\Bigl(1+\frac{r}{2}\Bigr)(S-s)\,,
   \label{E:omegaS1}
\end{equation}
\begin{eqnarray}
   \frac{\delta\omega_{\sigma\tau}(k)}{J}
   &=&\frac{4}{N}
       \biggl\{
         \biggl(\sqrt{\frac{S}{s}}+\sqrt{\frac{s}{S}}\biggr)
          \Bigl(
           \psi_{i(\sigma,\tau)}^{(2)}(k)
           \psi_{i(\sigma,\tau)}^{(4)}(k)
           {\rm cos}\frac{k}{2}
          +\frac{r}{2}
           \psi_{i(\sigma,\tau)}^{(1)}(k)
           \psi_{i(\sigma,\tau)}^{(2)}(k)
          \Bigr)
   \nonumber \\
   &&-
         \Bigl(1+\frac{r}{2}\Bigr)
         \big(
          \psi_{i(\sigma,\tau)}^{(2)}(k)^2
         +\psi_{i(\sigma,\tau)}^{(4)}(k)^2
         \big)
      \biggr\}
      \sum_{k'}
      \big(
       \psi_1^{(2)}(k')^2+\psi_2^{(2)}(k')^2
      \big)
   \nonumber \\
   &+&\frac{4}{N}
       \biggl\{
          \sqrt{\frac{S}{s}}\psi_{i(\sigma,\tau)}^{(2)}(k)^2
         +\sqrt{\frac{s}{S}}\psi_{i(\sigma,\tau)}^{(4)}(k)^2
       -2\psi_{i(\sigma,\tau)}^{(2)}(k)
         \psi_{i(\sigma,\tau)}^{(4)}(k)
         {\rm cos}\frac{k}{2}
      \biggr\}
   \nonumber \\
   &&\times
      \sum_{k'}
      \big(
       \psi_1^{(2)}(k')\psi_1^{(4)}(k')
      +\psi_2^{(2)}(k')\psi_2^{(4)}(k')
      \big)
      {\rm cos}\frac{k'}{2}
   \nonumber \\
   &+&\frac{2r}{N}
        \biggl\{
          \sqrt{\frac{S}{s}}\psi_{i(\sigma,\tau)}^{(2)}(k)^2
         +\sqrt{\frac{s}{S}}\psi_{i(\sigma,\tau)}^{(4)}(k)^2
       -2\psi_{i(\sigma,\tau)}^{(1)}(k)
         \psi_{i(\sigma,\tau)}^{(2)}(k)
      \biggr\}
   \nonumber \\
   &&\times
      \sum_{k'}
      \big(
       \psi_1^{(1)}(k')\psi_1^{(2)}(k')
      +\psi_2^{(1)}(k')\psi_2^{(2)}(k')
      \big)\,,
   \label{E:omegaS0}
\end{eqnarray}
with
\begin{eqnarray}
   &&
   \psi_{i(\sigma,\tau)}^{(1)}(k)
    =\frac{\sigma\tau}{R_i}
     \biggl\{
      \sqrt{\Bigl(1+\frac{r}{2}\Bigr)^2(S+s)^2
           -4Ss\Bigl({\rm cos}\frac{k}{2}-\tau\frac{r}{2}\Bigr)^2}
     -\sigma\Bigl(1+\frac{r}{2}\Bigr)(S+s)
     \biggr\}\,,
   \nonumber \\
   &&
   \psi_{i(\sigma,\tau)}^{(2)}(k)
    =\frac{2}{R_i}
     \Bigl({\rm cos}\frac{k}{2}-\tau\frac{r}{2}\Bigr)
     \sqrt{Ss}\,,
   \nonumber \\
   &&
   \psi_{i(\sigma,\tau)}^{(3)}(k)
    =-\frac{2\tau}{R_i}
     \Bigl({\rm cos}\frac{k}{2}-\tau\frac{r}{2}\Bigr)
     \sqrt{Ss}\,,
   \nonumber \\
   &&
   \psi_{i(\sigma,\tau)}^{(4)}(k)
    =-\frac{\sigma}{R_i}
     \biggl\{
      \sqrt{\Bigl(1+\frac{r}{2}\Bigr)^2(S+s)^2
           -4Ss\Bigl({\rm cos}\frac{k}{2}-\tau\frac{r}{2}\Bigr)^2}
     -\sigma\Bigl(1+\frac{r}{2}\Bigr)(S+s)
     \biggr\}\,,
   \nonumber \\
   &&
   R_{i(\sigma,\tau)}^2
    = 8\sigma\Bigl({\rm cos}\frac{k}{2}-\tau\frac{r}{2}\Bigr)^2Ss
     -2\sigma
      \biggl\{
       \sqrt{\Bigl(1+\frac{r}{2}\Bigr)^2(S+s)^2
            -4Ss\Bigl({\rm cos}\frac{k}{2}-\tau\frac{r}{2}\Bigr)^2}
      -\sigma\Bigl(1+\frac{r}{2}\Bigr)(S+s)
      \biggr\}^2\,.
\end{eqnarray}
We can obtain the dispersions (\ref{E:omegaS1}) by setting $k_y$
equal to $0$ and replacing $k_x$ and $r$ by $k$ and $r/2$,
respectively, in Eq. (\ref{E:omegab}).
The Bogoliubov transformation is given by Eq. (\ref{E:BT}) with
\begin{equation}
   {\cal U}
    =\left[
      \begin{array}{cccccccc}
       \psi_1^{(1)} & \psi_2^{(1)} & 0 & 0 &
       0 & 0 &-\psi_3^{(1)} &-\psi_4^{(1)} \\
       0 & 0 & \psi_3^{(2)} & \psi_4^{(2)} &
      -\psi_1^{(2)} &-\psi_2^{(2)} & 0 & 0 \\
       0 & 0 & \psi_3^{(3)} & \psi_4^{(3)} &
      -\psi_1^{(3)} &-\psi_2^{(3)} & 0 & 0 \\
       \psi_1^{(4)} & \psi_2^{(4)} & 0 & 0 &
       0 & 0 &-\psi_3^{(4)} &-\psi_4^{(4)} \\
       0 & 0 &-\psi_3^{(1)} &-\psi_4^{(1)} &
       \psi_1^{(1)} & \psi_2^{(1)} & 0 & 0 \\
      -\psi_1^{(2)} &-\psi_2^{(2)} & 0 & 0 &
       0 & 0 & \psi_3^{(2)} & \psi_4^{(2)} \\
      -\psi_1^{(3)} &-\psi_2^{(3)} & 0 & 0 &
       0 & 0 & \psi_3^{(3)} & \psi_4^{(3)} \\
       0 & 0 &-\psi_3^{(4)} &-\psi_4^{(4)} &
       \psi_1^{(4)} & \psi_2^{(4)} & 0 & 0 \\
      \end{array}
     \right]\,.
\end{equation}
\end{appendix}
\narrowtext

\begin{table}
\caption{The noninteracting-spin-wave (LSW) and interacting-spin-wave
         (ISW) estimates of the ground-state energy per rung,
         $E_{\rm g}/LJ$, for the $(S,s)=(1,\frac{1}{2})$ two-leg
         ferrimagnetic ladders, in comparison with the exact values
         (Exact) obtained by numerical diagonalization.
         The classical values (N\'eel), that is, the energies for the
         N\'eel configuration, are also listed for reference.}
\begin{tabular}{ccccc}
$r\equiv J'/J$ & N\'eel & LSW & ISW & Exact \\
\hline
$0.00$ & $-1.00$ & $-1.4365$ & $-1.4608$ & $-1.4542(1)$ \\
$0.01$ & $-1.01$ & $-1.4371$ & $-1.4638$ & $-1.4566(1)$ \\
$0.10$ & $-1.10$ & $-1.4539$ & $-1.4869$ & $-1.4807(1)$ \\
$0.50$ & $-1.50$ & $-1.6270$ & $-1.6478$ & $-1.6459(1)$ \\
$1.00$ & $-2.00$ & $-1.9295$ & $-1.9463$ & $-1.9433(1)$ \\
\end{tabular}
\label{T:Eg}
\end{table}

\widetext
\end{document}